\documentclass[letter,twocolumn]{jpsj3} 
\usepackage{color}
\title
{Nonadiabatic Dynamics of Ultracold Fermions in Optical Superlattices}

\author
{
Atsushi YAMAMOTO,\thanks{E-mail address: yamamoto@tp.ap.eng.osaka-u.ac.jp}$^1$ Makoto YAMASHITA,$^{2,3}$ and Norio KAWAKAMI$^4$
}

\inst{
$^1$Department of Applied Physics, Osaka University, Suita, Osaka 565-0871, Japan\\
$^2$NTT Basic Research Laboratories, NTT Corporation, Atsugi, Kanagawa 243-0198, Japan\\
$^3$Japan Science and Technology Agency, CREST, Chiyoda-ku, Tokyo 102-0075, Japan\\
$^4$Department of Physics, Kyoto University, Kyoto 606-8502, Japan
}

\abst{
We study the time-dependent dynamical properties of two-component ultracold fermions in a one-dimensional optical superlattice by applying the adaptive time-dependent density matrix renormalization group to a repulsive Hubbard model with an alternating superlattice potential. We clarify how the time evolution of local quantities occurs when the superlattice potential is suddenly changed to a normal one. For a Mott-type insulating state at quarter filling, the time evolution exhibits a profile similar to that expected for bosonic atoms, where correlation effects are less important. On the other hand, for a band-type insulating state at half filling, the strong repulsive interaction induces an unusual pairing of fermions, resulting in some striking properties in time evolution, such as a paired fermion co-tunneling process and the suppression of local spin moments. We further address the effect of a confining potential, which causes spatial confinement of the paired fermions.
}

\kword{optical lattice, ultracold atoms, quantum dynamics, density matrix renormalization group}

\begin{document}
\maketitle

Ultracold atoms in optical lattices offer unprecedented potential for exploring 
many-body effects \cite{Rev}. This system is considered to be an ideal realization 
of the Hubbard model that has been intensively studied for many years in condensed matter physics. 
Recent successful demonstrations of the Mott-insulator phase transition \cite{Gre,Jor,Sch} 
prove that we can obtain a comprehensive understanding of complicated 
many-body effects from a quantitative comparison of experiments and theories.  

High controllability of experimental parameters is one of the outstanding features of ultracold atoms. 
In optical lattices, this allows us to study the nonadiabatic dynamics of ultracold atoms by measuring the time evolution 
of atomic states after a quick modification of the lattice potentials or a sudden change in inter-atomic interactions 
via Feshbach resonance.  
Experiments are carried out in an ultrahigh vacuum so that 
energy is conserved and the quantum dynamics can remain coherent during the time evolution. 
Various intriguing properties have already been reported: 
the collapse and revival of a Bose-Einstein condensate \cite{Gre2}, 
the quantum version of Newton's cradle \cite{Kin}, 
the co-tunneling of bound atom pairs \cite{Win}, 
spin dynamics mediated by exchange interactions \cite{And}, 
and second-order atom tunneling \cite{Fol}. 
These experimental observations have stimulated theoretical investigations. 
Several numerical approaches have been applied to the Bose-Hubbrad model \cite{Kol,Cra,Fle,Lig,Dal} 
and the Fermi-Hubbard model \cite{Kol2,Tre} to reveal the nonequilibrium dynamics 
of many-body states that is largely unavailable in traditional condensed matter systems. 
However, most theoretical investigations have focused on bosonic systems  
owing to the current experimental situation.
This naturally motivates us to study the nonadiabatic dynamics of fermions with strong correlations.

In this paper, we investigate the nonadiabatic dynamical properties of two types of insulating states in a one-dimensional (1D) optical superlattice with two-component correlated fermions. First, we show the time evolution of the insulating states when there is a sudden change in a superlattice potential. We clarify how the particle correlations emerge in the nonadiabatic process. In particular, we clarify that the unusual pairing of fermions caused by the strong repulsive interaction produces a characteristic time-dependent profile of the physical quantities. The effect of a harmonic confining potential is also addressed.

Let us start with an extended Hubbard model, which may describe two-component ultracold fermions in a 1D optical superlattice with two-site periodic potential, 
\begin{eqnarray}
{\cal H} &=&  -J \sum_{i=1, \sigma}^L \left( c_{i,\sigma}^{\dagger}
c_{i+1,\sigma} + H.c.\right) + U \sum_{i=1}^{L} n_{i,\uparrow}n_{i,\downarrow} \nonumber\\
& & \!\!\!\!\! - V_{d} \sum_{i=odd,\sigma} n_{i } +V_{c} \ \sum_{i=1}^L\left( i-\frac{L+1}{2}\right) ^2 n_{i},
\label{eq:model}
\end{eqnarray}
where $c^\dag_{i,\sigma}$ ($c_{i,\sigma}$) is a creation (annihilation) operator of a fermion at the $i$-th site with spin $\sigma(=\uparrow, \downarrow)$. Here $J$ is the nearest-neighbor hopping integral, $L$  the total number of lattice sites, $U$ $(>0)$ is the on-site repulsive interaction and $V_d$ is the strength of the two-site periodic superlattice potential. $V_c$ is the curvature of a harmonic confining potential, which will be dealt with later in this paper. The ground state properties of the system without a confining potential were clarified in connection with quasi-1D organic molecules \cite{Pen,Fab,Ots,Fuk}, and the effect of confinement was addressed in the context of cold fermions \cite{Yam}. Among several intriguing phases caused by the superlattice structure \cite{Pen,Fab,Ots,Fuk,Yam}, we here consider two types of insulating states as an initial state of time evolution: Mott-type and band-type insulating states, which are respectively realized at quarter- and half-filling with a large potential difference, $V_{d} \gg J$. 

\begin{figure}[tbp]
  \begin{center}
   \includegraphics[clip,width=7.5cm]{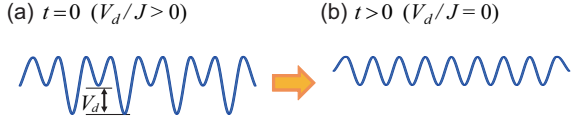}
    \caption{(Color online) 
    Schematic diagrams of nonadiabatic control of one-dimensional optical superlattice: 
    (a) an initial two-site periodic superlattice with a large potential difference $V_{d}$ and (b) a normal lattice after a sudden disappearance of $V_{d}$ for $t>$0.}
\label{fig0}
  \end{center}
\end{figure}

In the following, we deal with the dynamics of the above fermionic superlattice at quarter- and half-filling (we temporarily set $V_c=0$). To realize nonadiabatic control of our system, we suddenly change the lattice potential from $V_{d}/J=20 (t=0)$ to $V_{d}=0(t>0)$ as shown in Fig.\,\ref{fig0}. We make use of the adaptive time-dependent density matrix renormalization group (adaptive t-DMRG) \cite{Whi} method which is an extended version of the original DMRG \cite{Whi2} supplemented by the time-evolving block decimation  algorithm \cite{Vid}. The adaptive t-DRMG method can deal with the real-time dynamics of strongly correlated systems. We thus calculate the time-dependence of the density profile $\langle n_{i} \rangle$, the variance of local spins $ \Delta S^{z}_{i}$ (=$ \langle {S^{z}_{i}}^2\rangle$-$\langle S^{z}_{i} \rangle^2$) and the  double occupancy of fermions $\langle n_{i,\uparrow} n_{i,\downarrow} \rangle$. We check the accuracy of t-DMRG by comparing the results with the time-dependent exact diagonalization for small systems. In the following, we adopt $J$ as units of energy and time is measured in the unit of ${1}/{\hbar}$. 

\begin{figure}[tbp]
  \begin{center}
   \includegraphics[clip,width=8.0cm]{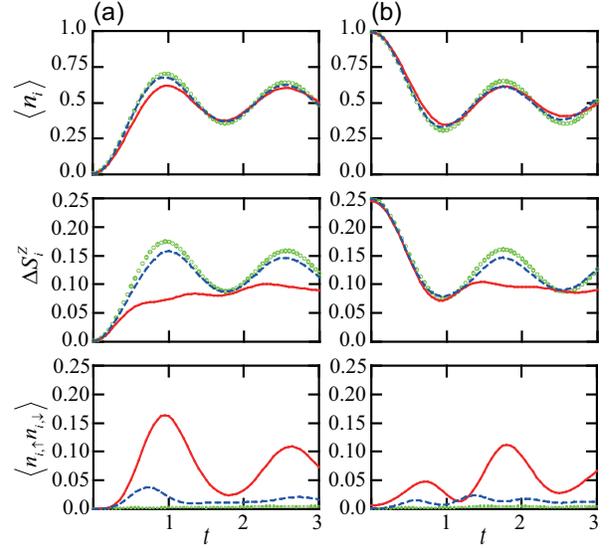}
    \caption{(Color online) Time dependence of the local quantities for several choices of on-site interaction for the system at quarter filling with $L=36$ and $N=18$. Plots of local density $\langle n_{i} \rangle$, variance of local spins $\Delta S^{z}_{i}$, and double occupancy $\langle n_{i,\uparrow} n_{i,\downarrow} \rangle$ from top to bottom. The on-site interaction is chosen as $U=1$ (solid line), $5$ (dashed line), and $9$ (dotted line) at (a) $i=18$ and (b) $i=19$.}\label{fig1}
  \end{center}
\end{figure}

First, we investigate the nonadiabatic time evolution of a Mott insulating state at quarter-filling. Fermionic atoms are initially located at every other site because of the large potential difference $V_d$(=20) between even and odd sites, which may be described by a sequence of empty and singly occupied sites 0,1,0,1, $\ldots$0.  Thus the system is a variant of a Mott insulator realized in a superlattice \cite{Pen,Fab,Ots,Fuk,Yam}. In Fig.\,\ref{fig1}, we show the time evolution of local quantities after $V_d$ has been suddenly turned off.
In Fig.\,\ref{fig1}(a)  (Fig.\,\ref{fig1}(b)) we show the time-dependent local density $\langle n_{i} \rangle$ at a given site $i=18$ ($i=19$) that is unoccupied (singly occupied) initially at $t=0$.
We find that the time-evolution profile is little affected by the interaction, and its oscillation period and amplitude are similar to those observed for bosonic atoms (approximately given by the Bessel function of the first kind) \cite{Cra,Fle}.
Note that $\langle n_{i} \rangle$ at even and odd sites always satisfies the conservation law of total fermions: the sum of $\langle n_{i} \rangle$ in the unit cell gives unity.  Although particle correlations do not appear to be very important for $\langle n_{i} \rangle$, they are somewhat relevant as regards the variance of local spins $\Delta S^{z}_{i}$ and the double occupancy $\langle n_{i,\uparrow} n_{i,\downarrow} \rangle$.
Figure\,\ref{fig1} indeed shows that the double occupancy is suppressed and its oscillation period becomes shorter as $U$ increases. Moreover, the double occupancy evolves differently at odd and even
sites (Fig.\,\ref{fig1}(a) and (b)). Let us first look at the even site in Fig.\,\ref{fig1}(a). For a small $U$, fermions move rather freely, so that $\langle n_{i,\uparrow} n_{i,\downarrow} \rangle$ is roughly given by the square of $\langle n_{i} \rangle$. However, for a larger $U$, the system needs the additional energy $U$ to realize double occupancy at even sites, which thus reduces its probability and shortens the oscillation period (see discussions below for the oscillation period).
For odd sites, as seen in Fig.\,\ref{fig1}(b), the amplitude of $\langle n_{i,\uparrow} n_{i,\downarrow} \rangle$ is smaller than that for even sites, since doubly occupied states at odd sites may be dominated by higher order hopping processes.
The variance of local spin fluctuations $ \Delta S^{z}_{i}$ shows somewhat complicated behavior in the intermediate $U$ region, since it is given by $\frac{1}{4}(\langle n_{i} \rangle - 2 \langle n_{i,\uparrow} n_{i,\downarrow} \rangle)$.
With a large $U$ limit, however, $ \Delta S^{z}_{i}$ is simply proportional to the local density $\langle n_{i} \rangle$, since the double occupancy tends to be zero.

\begin{figure}[tbp]
  \begin{center}
   \includegraphics[clip,width=8.0cm]{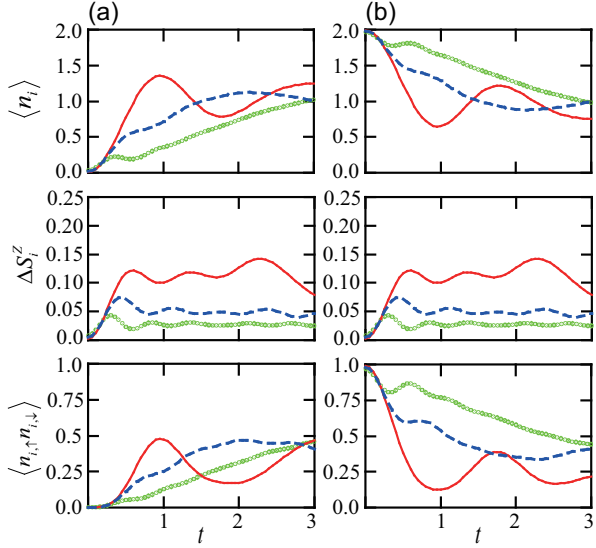}
    \caption{(Color online) Similar plots of the local quantities to those in Fig.\ref{fig1}: half-filling case with $L=36$ and $N=36$. 
The on-site interaction is chosen as $U=1$ (solid line), $5$ (dashed line), and $9$ (dotted line) at (a) $i=18$ and (b) $i=19$.}
\label{fig2}
  \end{center}
\end{figure}

We next investigate the time evolution of a band insulating state realized at half filling.
Owing to the two-site periodic superlattice structure, the atoms are initially distributed with a spatial configuration of 0,2,0,2, $\ldots$,0 since $V_d$(=20) is very large.
 This drives the system to a kind of band insulating state \cite{Pen,Fab,Ots,Fuk,Yam}.
To elucidate the dynamical properties of the band insulating state at half filling, we study the time evolution of the physical quantities mentioned above.
In Fig.\,\ref{fig2}, we present results computed for several different choices of $U$.
Let us first focus on the local density $\langle n_{i} \rangle$ for even and odd lattice sites as shown in Fig.\,\ref{fig2}(a) and (b).
The local density at even and odd sites clearly exhibits the particle-hole symmetry.
It is seen that the oscillation period decreases as $U$  increases.
The short time region shows the $t^2$-dependence and its coefficient does not depend on $U$, which is also the case for quarter filling.
On the other hand, the oscillation period in $\langle n_{i} \rangle$ is strongly affected by the on-site interaction, in contrast to the quarter filling case.
The period is approximately estimated in the limit of $V_{d} \gg U \gg J, t$, where the local density can be expressed as $\frac{4J^{2}}{U^2}(1- \cos (Ut))$ based on time-dependent perturbation theory.
This implies that the oscillation period is changed as $1/U$, which is indeed in accordance with the numerical results.
The double occupancy $\langle n_{i,\uparrow} n_{i,\downarrow} \rangle$ shows similar properties to the local density as regards its oscillating properties, although the former exhibits $t^4$-dependence in the short time region.
The variance of local spin fluctuations is given by $\Delta S^{z}_{i}=\frac{1}{4}(\langle n_{i} \rangle - 2 \langle n_{i,\uparrow} n_{i,\downarrow} \rangle)$, and therefore it exhibits the same oscillation periodicity as for the local density and the double occupancy.
Note that $\Delta S^{z}_{i}$ exhibits similar profiles for even and odd sites, owing to the particle-hole symmetry.

\begin{figure}[tbp]
  \begin{center}
   \includegraphics[clip,width=7.0cm]{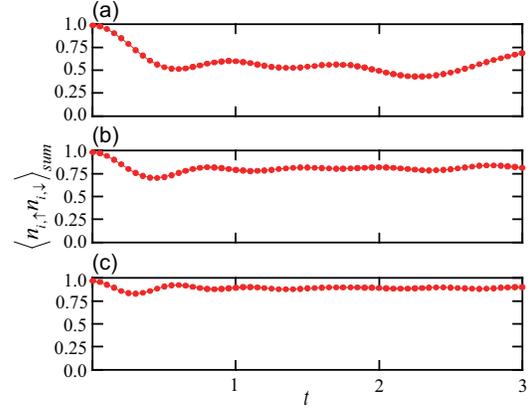}
    \caption{(Color online) Plots of double occupancy summed in the unit cell for a system with $L=36$ and $N=36$: (a) $U=$1.0, (b) $U=$5.0, and (c) $U=$9.0. }\label{fig3}
  \end{center}
\end{figure}

In addition to the characteristic oscillating properties, we encounter some striking behavior at half filling.
As seen in Fig.\,\ref{fig2}, the variance of spin fluctuations $\Delta S^{z}_{i}$ at $i=18$ and $i=19$ is suppressed as $U$ increases, in contrast to the naive expectation that the local spin develops with increasing $U$.
Also, all the quantities cease to oscillate as $U$ increases.
To clarify the origin of this behavior, we plot the double occupancy summed in the unit cell $\langle n_{i,\uparrow} n_{i,\downarrow} \rangle _{sum}$ $(= \langle n_{i=odd,\uparrow} n_{i=odd,\downarrow} \rangle$ + $ \langle n_{i=even,\uparrow} n_{i=even,\downarrow} \rangle$) in Fig.\,\ref{fig3}.
It is notable that $\langle n_{i,\uparrow} n_{i,\downarrow} \rangle _{sum}$  ($\simeq 0.9$) is hardly changed with time evolution for a large $U$, while it shows rather a strong time dependence for a small $U$.
With a sudden change in $V_d$, two atoms occupying an odd site try to move apart from each other, but this process is not allowed for $U \gg J$, since the large potential energy $U$ cannot be transferred to the kinetic energy.
This leads to an unusual pairing of fermions induced by a repulsive interaction.
The resulting bound pairs can hop around lattice sites via a higher tunneling process called the co-tunneling process.
For bosons, similar bound states and the associated co-tunneling process were proposed and observed experimentally \cite{Win}.
The bound pairs induced by the repulsive interaction are stable during time evolution and suppress the development of local spins, thus reducing $\Delta S^{z}_{i}$ for a large $U$.
This naturally explains the unusual suppression of $\Delta S^{z}_{i}$ found in Fig.\,\ref{fig2}, which is also confirmed by the fact that the local density profile is roughly given by twice the double occupancy $\langle n_{i,\uparrow} n_{i,\downarrow} \rangle$ for a larger $U$, as seen in Fig.\,\ref{fig2}. 

\begin{figure}[tbp]
  \begin{center}
   \includegraphics[clip,width=8.5cm]{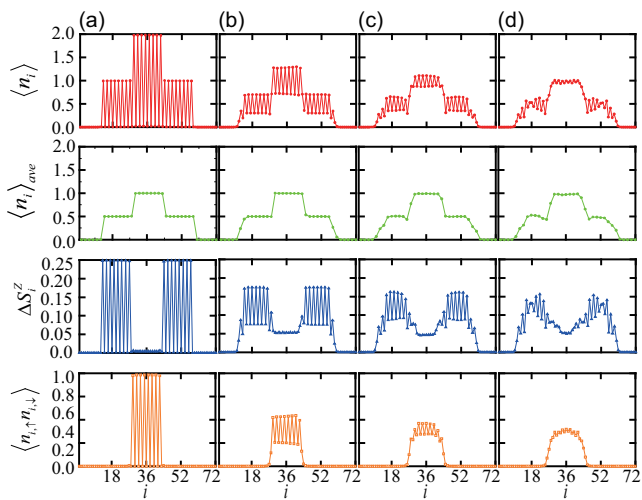}
    \caption{(Color online) Spatial distribution of the local fermion density $\langle n_{i} \rangle$, the average fermion density  $\langle n_{i} \rangle_{ave}$, the variance of local spins $\Delta S^{z}_{i}$, and the double occupancy $\langle n_{i,\uparrow} n_{i,\downarrow} \rangle$ for (a) $t=0$, (b) $t=1$, (c) $t=2$, and (d) $t=3$. The system parameters are taken as  $U=5.0$, $V_{c}=0.025$, $L=72$ and $N=32$. The initial state is in a coexisting phase including quarter- and half-filling insulating regions. Characteristic saw-toothed profiles are gradually smeared with time evolution.
}\label{fig4}
  \end{center}
\end{figure}

Finally, to make the model more accessible to experiments, we consider the effect of the harmonic confining potential given by the last term in Eq.(\ref{eq:model}).
Figure\,\ref{fig4} shows the time evolution of the physical quantities discussed above.
The initial ground state ($t$=0) is assumed to be in a coexisting phase including two different insulating regions that roughly satisfy the conditions of half- and quarter-filling. 
We have confirmed that all the quantities in each insulating region exhibit the characteristic time dependence discussed above: the two-site alternating profiles are gradually smeared after the superlattice potential $V_d$ is turned off.
According to the local density averaged in a unit cell $\langle n_{i} \rangle_{ave}$ $(=\frac{1}{2} (\langle n_{i=odd} \rangle + \langle n_{i=even} \rangle)$ shown in Fig.\,\ref{fig4}, we can see that the plateau structure is quite stable around the center (half-filling), while it gradually collapses around the quarter-filling regions since they are close to the edges of the system.
Although the spatial extent of the half-filling region ($\langle n_{i} \rangle_{ave}=1$) is little changed with time evolution, its insulating characteristic gradually disappears.
After the sudden potential change, the fermions form bound pairs and the resulting hard-core bosons can move via a co-tunneling process by keeping the number of doubly occupied sites unchanged.
Correspondingly, the spin is not developed even in the long-time region, although the system is in a good condition ($\langle n_{i} \rangle_{ave}=1$) to realize an ordinary Mott insulator for $t>0$.
On the other hand, the quarter filling regions are dominated by unpaired fermions with strong repulsion, which are characterized by large spin fluctuations and vanishing double occupancy, as seen in  Fig.\,\ref{fig4}. 

It should be stressed here that the half-filling region specified by $\langle n_{i} \rangle_{ave}=1$, which is rather robust during time evolution, is spatially confined around the center, because hard-core bosons (paired fermions) created around the center experience the high pressure imposed by strongly correlated unpaired fermions from outside.
This causes a unique confinement phenomenon as regards the paired fermions due to the strong repulsive interaction. 


In summary, by using adaptive t-DMRG simulations, we have uncovered several intriguing dynamical features of repulsive fermions in a 1D optical superlattice.
We have clarified how the correlated fermions evolve in time when a two-site periodic superlattice is suddenly changed into a normal one.
For a homogeneous system at quarter-filling, the time dependence exhibits a similar process to bosons.
On the other hand, in the half-filling case, more striking features emerge: the strong repulsive interaction induces fermion pairs, which can move around the lattice via a co-tunneling process.
This kind of pair formation, which is a fermionic analog of the bosonic case \cite{Win}, in turn suppresses the development of local spins.
We have clarified how this kind of pair hopping emerges via a crossover from single-particle motion in the weak-interaction limit.
We have also addressed the effect of the confining potential.
We have shown that for an initial state consisting of quarter- and half-filling insulating regions, an unusual confinement phenomenon involving paired fermions occurs with time evolution.
We expect that the nonadiabatic properties of correlated fermionic atoms described in this paper will be observed experimentally.

\acknowledgments
We thank Y.\,Fujihara, K.\,Inaba, S.\,Suga and S.\,Gurvitz for stimulating discussions.
This work was partly supported by the Grant-in-Aid for Scientific Research [No.20102008, 20029013, 21540359] and the Global COE Program ``The Next Generation of Physics, Spun from Universality and Emergence" from MEXT, Japan.
The numerical computations were carried out at the Supercomputer Center, the Institute for Solid State Physics, University of Tokyo.
A.\,Yamamoto is supported by JSPS Research Fellowships for Young Scientists.

%

\end{document}